\documentclass[aps,pra,twocolumn,superscriptaddress,showpacs]{revtex4}
\usepackage{amsmath}
\usepackage{amsfonts}
\usepackage{graphicx}
\newcommand{\be}{\nopagebreak[3]\begin{equation}}
\newcommand{\ee}{\end{equation}}
\newcommand{\ba}{\nopagebreak[3]\begin{eqnarray}}
\newcommand{\ea}{\end{eqnarray}}

\newcommand{\bmult}{\nopagebreak[3]\begin{multline}}
\newcommand{\emult}{\end{multline}}

\begin{document}
\title{Effects of Phase Fluctuations on Phase Sensitivity and Visibility\\
of Path-Entangled Photon Fock States}
\author{Bhaskar Roy Bardhan}
\email{broyba1@lsu.edu}
\author{Kebei Jiang}
\affiliation{Hearne Institute for Theoretical Physics and Department of Physics and Astronomy, Louisiana State University, Baton Rouge, LA 70803}
\author{Jonathan P.\ Dowling}
\affiliation{Hearne Institute for Theoretical Physics and Department of Physics and Astronomy, Louisiana State University, Baton Rouge, LA 70803}
\affiliation{Computational Science Research Center, Beijing, China 100084}

\date{\today }

\begin{abstract}

We study effects of phase fluctuations on phase sensitivity and visibility of a class of robust path-entangled photon Fock states (known as $mm^{\prime}$ states) as compared to the maximally path-entangled N00N states in presence of realistic phase fluctuations such as turbulence noise. Our results demonstrate that the $mm^{\prime}$ states, which are more robust than the N00N state against photon loss, perform equally well when subject to such fluctuations. We derive the quantum Fisher information with the phase-fluctuation noise, and show that the phase sensitivity with parity detection for both of the above states saturates the quantum Cram\'er-Rao bound in presence of such noise, suggesting that the parity detection presents an optimal detection strategy.

\end{abstract}


\maketitle

\section{Introduction}

Quantum states of light such as squeezed sates or entangled states, have long been known to produce greater precision, resolution and sensitivity in metrology, imaging, and object ranging~\cite{Caves,JPD,Kapale,Giovannetti} than what is possible classically.
One of the most prominent examples of such a non-classical state is the N00N state~\cite{Boto,Kok,Sanders}, which is an equal coherent superposition of $N$ photons in one path of a Mach-Zehnder interferometer with none in the other, and vice-versa. This state may be written as $| N::0 \rangle=(| N, 0 \rangle + | 0, N \rangle)/\sqrt{2}$, and can be used to achieve Heisenberg-limited supersensitivity as well as super resolution in quantum metrology~\cite{Boto,Durkin}. In recent years, several schemes for reliable production of such states have been proposed, making them useful in super-precision measurements in optical interferometry, atomic spectroscopy, gravitational wave detection, and magnetometry along with potential applications in rapidly evolving fields such as quantum imaging and sensing~\cite{Mitchell,Walther,Nielsen,Nagata,Vitelli,Spagnolo,Jones2}.

However, due to inevitable interactions with the surrounding environment, the N00N state tends to decohere in presence of noisy environment.  Recently, a few authors investigated the effects of photon loss on the performance of N00N state in quantum interferometric setups~\cite{Gilbert, Rubin, Parks, Huver, Kebei}, that demonstrate that N00N states undergoing loss decohere very rapidly, making it difficult to achieve super-sensitivity and resolution in a lossy environment. Huver \emph{et al.} proposed a class of generalized Fock states, known as $mm^{\prime}$ states, by introducing decoy photons to the N00N state in both paths of the interferometer, and showed that such states provide better metrological performance than N00N states in presence of photon loss~\cite{Huver}.

In real life applications such as a quantum sensor or radar,  phase fluctuation due to different noise sources can further degrade the phase-sensitivity by adding significant noise to the phase $\phi$ to be estimated or detected. For instance, when one considers propagation of the entangled states over distances of kilometers, through say the atmosphere, then atmosphere turbulence becomes an issue as it can cause uncontrollable noise or fluctuation in the phase. In this sense, phase-fluctuation stands as the most detrimental for phase estimation, rendering the quantum metrological advantage for achieving super-sensitivity  and super-resolution totally useless. It is therefore imperative to investigate the impacts of such random phase-fluctuations on the phase-sensitivity of quantum mechanically entangled states. In particular, we consider both the $mm^{\prime}$ and N00N states, and show how the phase-sensitivity and visibility of the phase signal are affected by added phase-fluctuations caused by turbulence noise.

We study the parity detection~\cite{Gerry} for the interferometry with the phase-fluctuated $mm^{\prime}$ and N00N states. This detection scheme has been shown to reach Heisenberg limited sensitivity when combined with the lossless N00N state~\cite{Gerry, Campos, Gerry-Mimih,Kaushik}. Here we calculate the minimum detectable phase shift in presence of the turbulence noise, and show that the lower bound of the phase-fluctuated sensitivity for both the states saturates the quantum Cram\'er-Rao bound~\cite{Braunstein1,Braunstein2}, which gives the ultimate limit to the precision of the phase measurement. This result suggests that the parity detection serves as an optimal detection strategy when the given states are subject to the phase-fluctuations.

The paper is organized as follows. In section II, we introduce the $mm^{\prime}$ and N00N states, and describe their evolution under the phase fluctuations. We define the parity detection operator in section III, and calculate the phase-sensitivity and the visibility with the phase noise using the parity operator.
In section IV, we derive lowest possible uncertainty (quantum Cram\'er-Rao bound) in estimating the phase $\phi$ for these path-entangled Fock states, and show that the parity operator saturates the quantum Cram\'er-Rao bound for both $mm^{\prime}$ and N00N states. Using the same detection technique, we then derive the phase-sensitivity and the visibility in a more general case with both the photon loss and phase-fluctuations in Sec V. Section VI contains our concluding remarks, and further outlook with the potential implementations of the phase estimation with fluctuating phase noise.

\section{Evolution of the $mm^{\prime}$ and N00N states under phase fluctuations}

The states we now wish to investigate are the following: 
\begin{align}
|m::m^\prime \rangle_{a,b} = \frac{1}{\sqrt{2}}(|m,m^\prime \rangle_{a,b} + |m^\prime, m\rangle_{a,b})
\end{align}
where $a$ and $b$ indicate the two paths of a two-mode optical interferometer. These states are called the $mm^{\prime}$ states, and they can be produced, for example, by post-selecting on the output of a pair of optical parametric oscillators~\cite{Glasser}. 

The $mm^{\prime}$ state reduces to a N00N state when $m=N$ and $m^\prime = 0$, leading to
\begin{align}
|N::0\rangle_{a,b} = \frac{1}{\sqrt{2}}(|N,0\rangle_{a,b} + |0,N\rangle_{a,b})
\end{align}
The $mm^{\prime}$ states have been shown to be more robust than the N00N states against photon loss~\cite{Huver,Kebei}. In following calculations, we drop the subscripts and always assume the first number in a ket or bra corresponds to mode $a$ while the second to mode $b$.

We start with the propagation of $mm^{\prime}$ and N00N states through a simplified Mach-Zehnder interferometer as shown in Fig. 1, where details of source and detection are represented by their respective boxes. The input state at stage I is presented by Eq. (1) with the photon number difference $(\Delta m = m - m^{\prime})$ between the two arms is fixed.

\begin{figure}
\centering
	\includegraphics[width=\columnwidth]{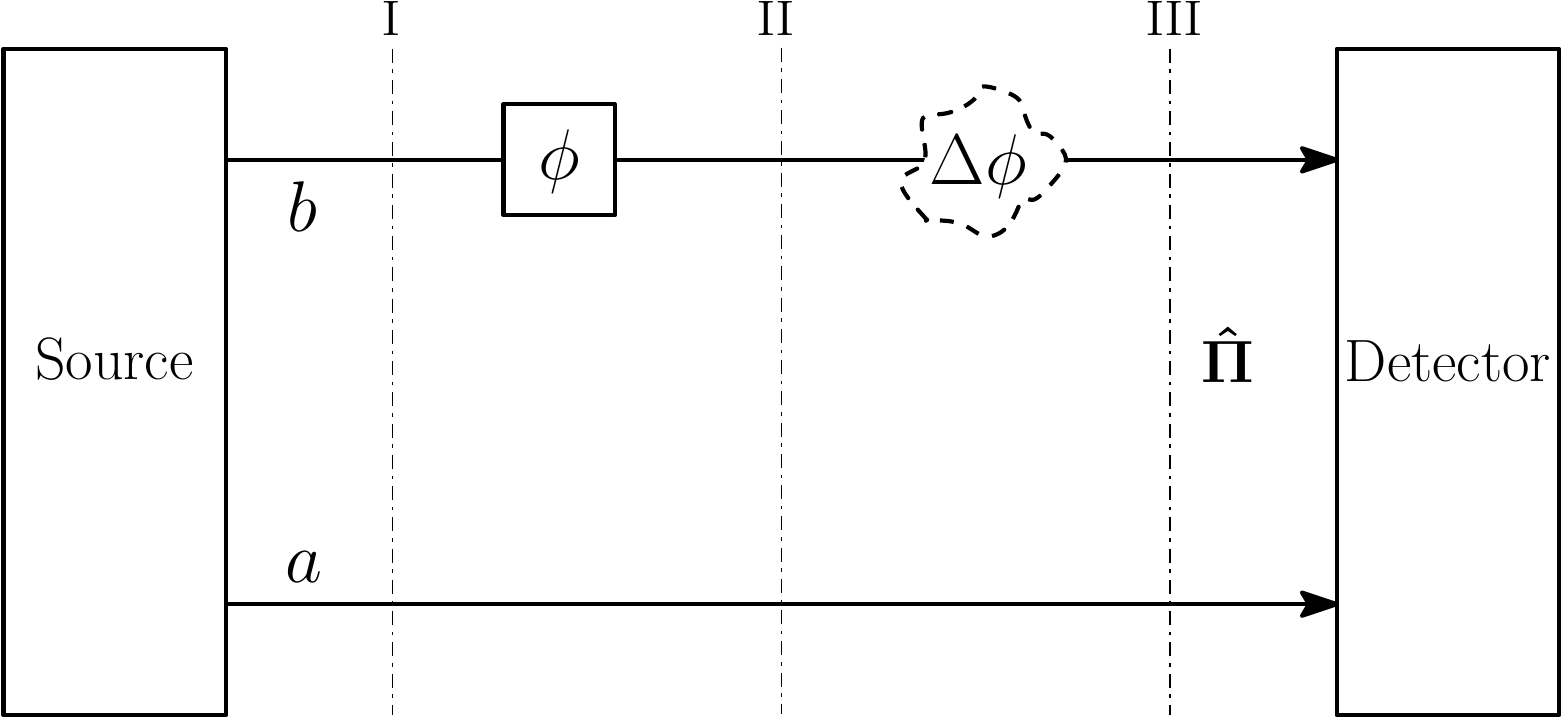}
	\caption{Schematic diagram of a simplified Mach-Zehnder interferometer with the modes $a$ and $b$ for the $mm^{\prime}$ and N00N states as the input. The source and detector in the interferometer are represented by the respective boxes. Effects of the phase-fluctuations due to the turbulence noise is represented by $\Delta \phi$ in the upper path b of the interferometer. The upper beam passes through a phase-shifter $\phi$, and the phase acquired depends on the total number of photons $\Delta m=m - m^{\prime}$ (or $N$) passing throughout the upper path. Transformed parity detection is used as the detection scheme at both of the two modes at stage III inside the interferometer.}
	\label{Fig1}
\end{figure}

The presence of the phase-shifter in the upper path $b$ introduces a phase-shift $\phi$ to the photons traveling through it, so that the state at stage II becomes
\begin{align}
|\psi\rangle_{\textrm{II}} =& \frac{1}{\sqrt{2}} (e^{im^{\prime} \phi} |m,m^\prime\rangle + e^{im \phi} |m^\prime, m\rangle) \nonumber\\
=& \alpha |m, m^\prime \rangle + \beta | m^\prime, m\rangle,
\end{align}
where $\alpha = e^{im^\prime \phi}/\sqrt{2} $ and $\beta = e^{im \phi}/\sqrt{2} $.
Because of the different number of photons being phase-shifted on the upper path $b$, phase shifts accumulated are different along the two paths, thus providing the possibility of interference upon detection.

The combined effects of random phase fluctuations are represented by  $\Delta \phi$ in the upper path in Fig. 1, and the $mm^{\prime}$ state at stage III is then given by,
\begin{align}
\vert \psi(\Delta \phi) \rangle_{\textrm{III}} = \alpha e^{im'\Delta \phi} |m, m^\prime\rangle + \beta e^{im\Delta \phi} | m^\prime, m\rangle.
\end{align}
Notice that because of the random nature of the phase fluctuations, the state of the system becomes a mixed state and the associated density matrix is then
\begin{align}
\rho_{mm'} &=\langle \vert \psi(\Delta \phi) \rangle_{\textrm{III}} ~_{\textrm{III}} \langle \psi (\Delta \phi) \vert \rangle.
\end{align}
Random fluctuations $\Delta \phi$ in the phase effectively causes the system to undergo pure dephasing. As a result, the off-diagonal terms in the density matrix will acquire decay terms, while the diagonal terms representing the population will remain intact, \emph{i.e.} the photon number will be preserved along the path~\cite{James}.

We can expand the exponential in Eq. (4) in a series expansion, and consider the terms up to the second order in $\Delta \phi$. We assume the random phase fluctuation $\Delta \phi$ to have Gaussian statistics described by Wiener process, \emph{i.e.} with zero mean and non-zero variance $\langle \Delta \phi^{2}\rangle=2 \Gamma L$ ($L$ is the length of the dephasing region, and $\Gamma$ is the dephasing rate). Ensemble averaging over all realizations of the random process then gives,
\begin{align*}
\langle e^{i \Delta m \Delta \phi} \rangle&=1+i \Delta m \langle \Delta \phi \rangle-(\Delta m)^{2} \langle \Delta \phi^{2} \rangle /2 \nonumber \\
&= 1-(\Delta m)^{2} \Gamma L \approx e^{-(\Delta m)^{2} \Gamma L}.
\end{align*}
The density matrix for the $mm^{\prime}$ state is given by
\begin{align}
\rho_{mm'}=~& |\alpha|^{2} |m,m' \rangle \langle m,m'| +|\beta|^{2} |m', m\rangle \langle m',m| \nonumber \\
~&+\alpha^* \beta  e^{-(\Delta m)^2 \Gamma L} |m,m' \rangle \langle m',m|  \nonumber \\
~&+ \alpha \beta^* e^{-(\Delta m)^2 \Gamma L} |m', m \rangle \langle m,m'|. 
\end{align}
This result agrees with the density matrix obtained from solving the master equation in Ref.~\cite{James}. The similar equation for the N00N state can be obtained from Eq. (2) as
\begin{align}
\nonumber
\rho_{\textrm{N00N}}&= |\alpha|^{2} |N,0 \rangle \langle N,0| +|\beta|^{2} |0,N \rangle \langle 0,N | \nonumber \\
~&+ \alpha^* \beta e^{-N^2 \Gamma L} |N,0 \rangle \langle 0,N|+~ \alpha \beta^* e^{-N^2 \Gamma L} |0,N \rangle \langle N,0|
\end{align}

\section{Parity Operator}
Achieving super-resolution and super-sensitivity depends not only on the state preparation, but also on the optimal detection schemes with specific properties.
In this paper, we study parity detection, which was originally proposed by Bollinger \emph{et al.} in the context of trapped ions~\cite{Bollinger} and it was later adopted for optical interferometry by Gerry~\cite{Gerry}. The original parity operator can be expressed as $\hat{\pi} = \exp(i\pi\hat{n})$, which distinguishes states with even and odd number of photons without having to know the full photon number counting statistics. 
Usually the parity detection is only applied to one of two output modes of the Mach-Zehnder interferometer. In our case,  the parity operator inside the interferometer, following Ref.~\cite{Aravind}, can be written as
\begin{align}
\hat{\Pi} = i^{(m+m^{\prime})} \sum_{k=0}^{m}(-1)^k |k,n -k \rangle\langle n-k, k|,
\end{align}
where $\hat{\Pi}^2 = 1$ and $n = m+m^\prime$, is the total number of photons. And it should be noticed that the parity operator inside the interferometer detects both mode $a$ and $b$ of the field.

The expectation value of the parity for the $mm^{\prime}$ state is then calculated as
\begin{align}
\langle \hat{\Pi}\rangle_{mm^\prime} &= \mathrm{Tr}\left[ \hat{\Pi}\rho_{mm^\prime} \right]\nonumber \\
&=(-1)^{(m+m^{\prime})}e^{-(\Delta m)^2 \Gamma L} \cos [\Delta m(\phi-\pi/2)],
\end{align}
where the density matrix $\rho_{mm^\prime}$ is given by Eq. (6). If we put a half-wave plate in front of the phase shifter, which amounts to replace $\phi$ by $\phi+\pi/2$, the expectation value becomes,
\begin{align}
\langle \hat{\Pi}\rangle_{mm^\prime} =& (-1)^{(m+m^{\prime})}e^{-(\Delta m)^2 \Gamma L} \cos[\Delta m \phi].
\end{align}
Using the density matrix $\rho_{\textrm{N00N}}$ in Eq. (7) for the N00N state, we can also obtain the expectation value of the parity operator for the N00N state as
\begin{align}
\langle \hat{\Pi} \rangle_{\textrm{N00N}} =&~ \mathrm{Tr}\left[\hat{\Pi}\rho_{\textrm{N00N}}\right]\nonumber \\ 
=&~ (-1)^{N} e^{-N^2 \Gamma L}  \cos [N\phi].
\end{align}

\subsection{Phase-sensitivity}

In quantum optical metrology, the precision of the phase measurement is given by the phase-sensitivity.
We now calculate the phase sensitivity for both the $mm^{\prime}$  and N00N states using the expectation values of the parity operator obtained above.

\begin{figure}
\centering
	\includegraphics[width=\columnwidth,height=2.4in]{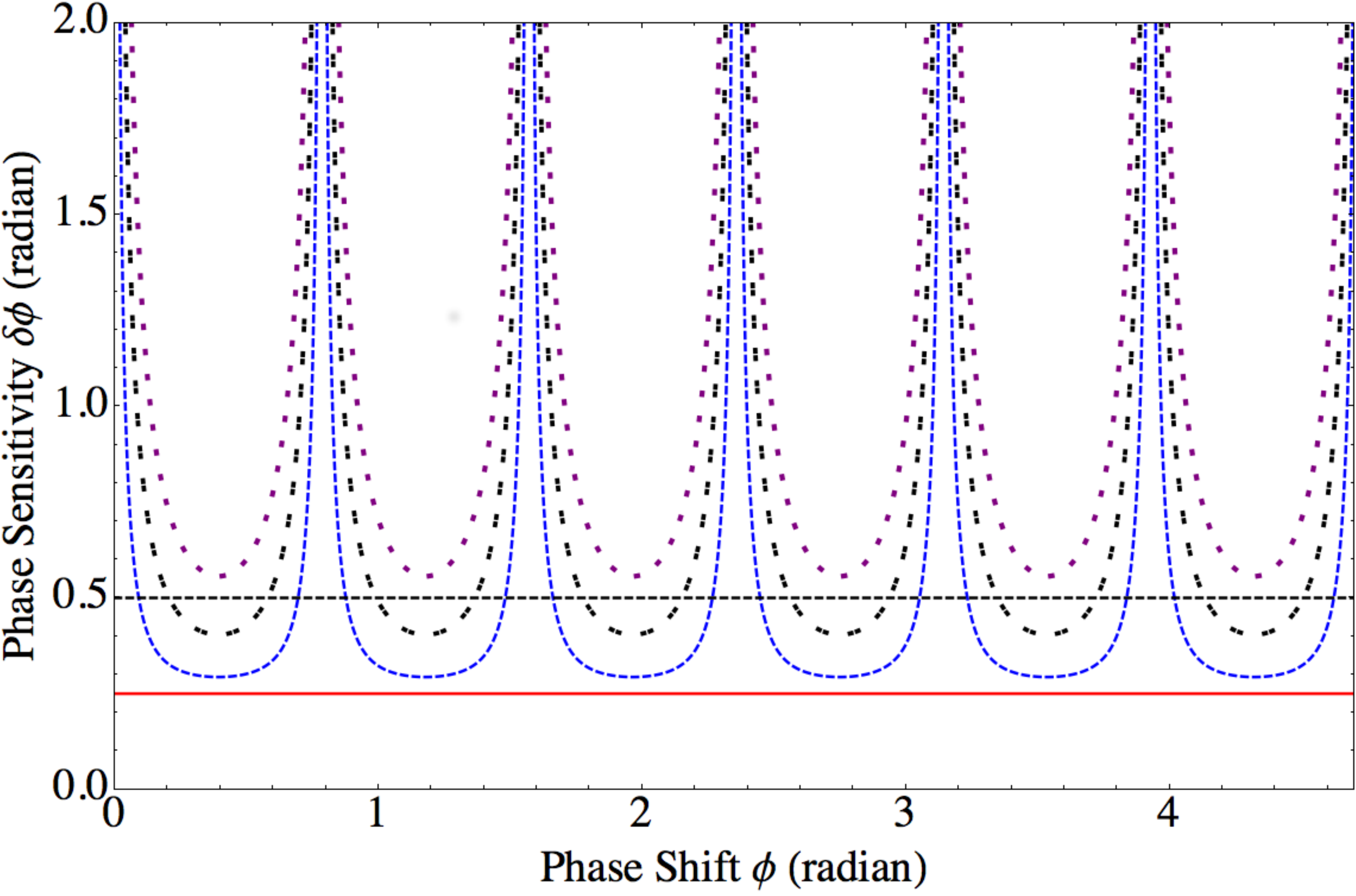}
	\caption{(Color online) Phase sensitivity $\delta \phi$ of the $mm^{\prime}$ state $|5::1 \rangle$), or the N00N state $|4::0 \rangle$, having the same phase information, as a function of phase shift $\phi$ from a two-mode interferometer for different values of $\Gamma$: $\Gamma=0.1$ (curved blue dashed line), $\Gamma=0.3$ (curved black double-dotted line), $\Gamma=0.5$ (curved purple dotted line). The Heisenberg limit ($1/N$) and the shot noise limit ($1/\sqrt{N}$) of the phase sensitivity for the N00N state are shown by the red solid line and the black dashed line, respectively, for comparison.}
\end{figure}

Phase sensitivity using the parity detection is defined by the linear error propagation method ~\cite{Bevington}
\begin{align}
\delta \phi =\frac{\Delta \hat{\Pi}}{\vert \partial \langle \hat{\Pi} \rangle / \partial \phi \vert},
\end{align}
where $\Delta \hat{\Pi}=\sqrt{\langle \hat{\Pi}^2 \rangle - \langle \hat{\Pi} \rangle^2}$. Given $\langle \hat{\Pi}_{mm'}^2 \rangle=1$ the phase-sensitivity with the parity detection for the $mm^{\prime}$ state is 
\begin{align}
\delta \phi_{mm^\prime} = \sqrt{\frac{1-e^{-2(\Delta m)^2 \Gamma L} \cos^2 (\Delta m \phi)}{(\Delta m)^2 e^{-2(\Delta m)^2 \Gamma L } \sin^2 (\Delta m \phi)}}.
\end{align}

\begin{figure}[h!]
\centering
	\includegraphics[width=\columnwidth]{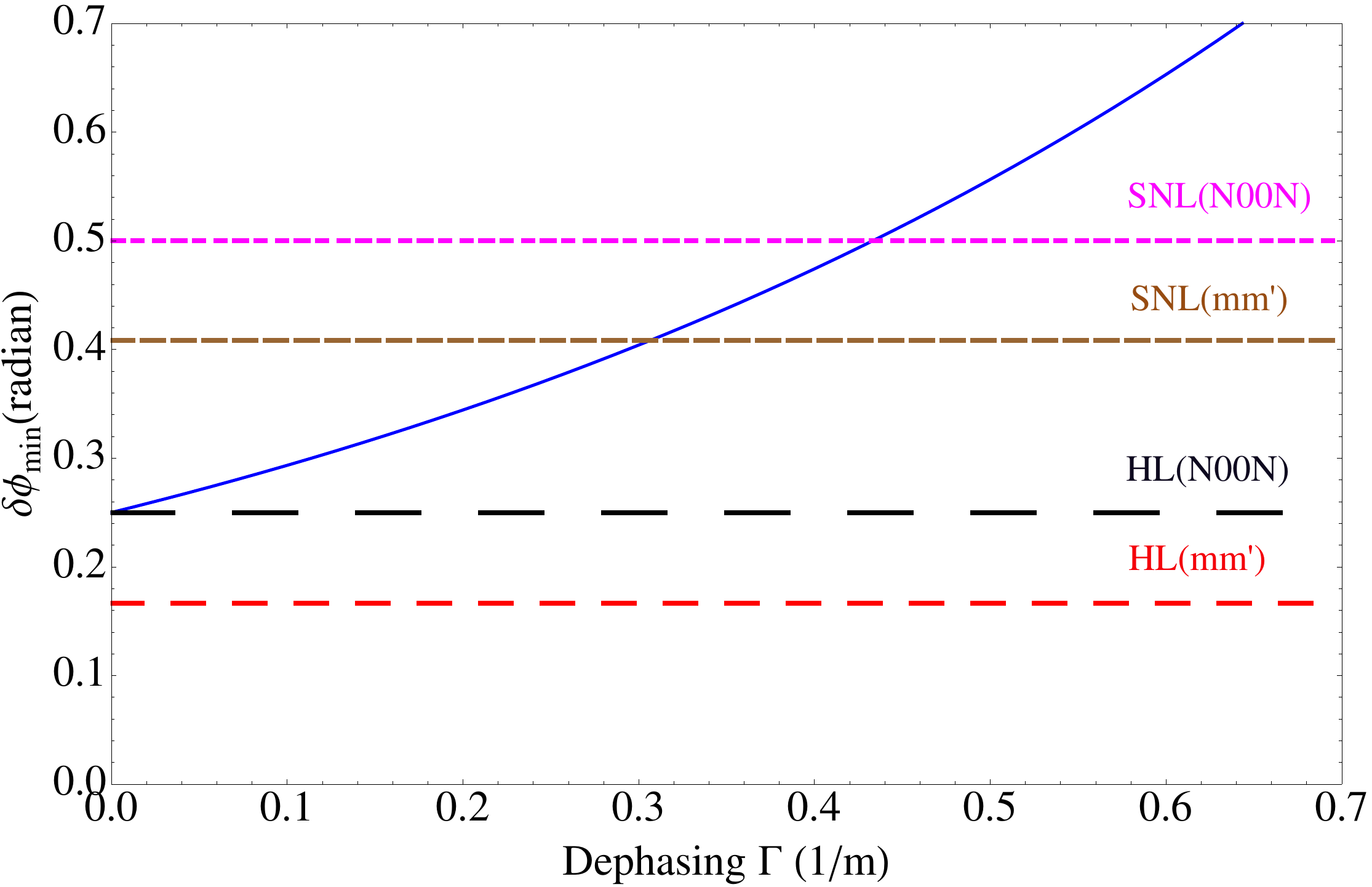}
	\caption{(Color online) Minimum phase-sensitivity $\delta \phi_{\textrm{min}}$ of the $mm^{\prime}$ state $|5::1 \rangle$), or the N00N state $|4::0 \rangle$ as a function of  $\Gamma$. The shot noise limits (SNL) and the Heisenberg limits (HL) of the phase-sensitivity for both the states are also shown for comparison.}
	\label{Fig3}
\end{figure}
For the N00N state the phase-sensitivity with the parity detection is similarly obtained as 
\begin{align}
\delta \phi_{\textrm{N00N}} =\sqrt{\frac{1-e^{-2N^2 \Gamma L} \cos^2 N \phi}{N^2 e^{-2N^2 \Gamma L } \sin^2 N \phi}}.
\end{align}
We note that in the limit of no dephasing $(\Gamma \rightarrow 0)$, $\delta \phi_{mm^\prime} \rightarrow 1/(\Delta m)$. For the N00N state, $\Gamma \rightarrow 0$ case leads to $\delta \phi_{\textrm{N00N}} \rightarrow 1/N$ (Heisenberg limit of the phase-sensitivity for the NOON state).

In Fig.~2, we plot the phase sensitivities $\delta \phi_{mm^\prime} $ and $\delta \phi_{\textrm{N00N}}$ for the various dephasing rates $\Gamma$ choosing $\Delta m= N$, so that the amount of phase information is the same for either state. For $\Delta m= N$, Eqs. (13) and (14) show that the $mm^{\prime}$ and N00N states give rise to the same phase-sensitivity. In particular, we show the phase-sensitivity for the states $|4::0 \rangle$ and $|5::1\rangle$, and find that both the states perform equally well in presence of phase fluctuations when parity detection is used, although the former has been shown to outperform N00N states in presence of photon loss~\cite{Huver,Kebei}.

The minimum phase-sensitivities $\delta \phi_{\textrm{min}}$ can be obtained from Eqs.~(13) and (14) for $\phi = \pi/(2 \Delta m)$, or $\phi = \pi/(2 N)$ for the $mm^{\prime}$ or N00N states, respectively. For the $|4::0 \rangle$ and $|5::1\rangle$ states, we plot the minimum phase-sensitivity $\delta \phi_{\textrm{min}}$ in Fig.~3 for as a function of $\Gamma$, and compare with the SNL and HL for both the states. 

The HL for a general $mm^{\prime}$ state is $1/(m+m')$ in terms of the total number of photons available, and is equal to $1/N$ for the N00N state. The SNL for these two states is given by $1/(\sqrt{m+m^{\prime}})$ and $1/\sqrt{N}$, respectively. In Fig.~3, we see that the minimum phase-sensitivity $\delta \phi_{\textrm{min}}$ hits the HL for the NOON state for $\Gamma=0$ only, while it never reaches the HL for the $mm^{\prime}$ state. However, $\delta \phi_{\textrm{min}}$ is below the SNL for both of the states for small values of $\Gamma$, but increase in the phase-fluctuation, \emph{i.e.} $\Gamma$, leads to the phase-sensitivity above the SNL, as shown in Fig.~3.

\subsection{Visibility}
We use the parity operator for the detection, and to quantify the degree of measured phase information we define the relative visibility as
\begin{align}
V_{mm^\prime} = \frac{\langle \hat{\Pi}_{mm^{\prime}}\rangle_{\textrm{max}} - \langle \hat{\Pi}_{mm^\prime}\rangle_{\textrm{min}}}{\langle \hat{\Pi}_{mm^\prime}(\Gamma=0)\rangle_{\textrm{max}} - \langle \hat{\Pi}_{mm^\prime}(\Gamma=0)\rangle_{\textrm{min}}},
\end{align}
where the numerator corresponds to the difference in the maximum and minimum parity signal in presence of phase fluctuations, while the denominator corresponds to the one with no dephasing, $i.e.$ $\Gamma= 0$. Visibility for the N00N state is similarly defined as 
\begin{align}
V_{\textrm{N00N}} =\frac{\langle \hat{\Pi}_{\textrm{N00N}}\rangle_\textrm{{max}} - \langle \hat{\Pi}_{\textrm{N00N}}\rangle_{\textrm{min}}}{\langle \hat{\Pi}_{\textrm{N00N}}(\Gamma=0)\rangle_{\textrm{max}} - \langle \hat{\Pi}_{\textrm{N00N}}(\Gamma=0)\rangle_{\textrm{min}}}
\end{align}

Using Eqs. (10) and (11), we then obtain the visibilities for the $mm^{\prime}$ state
\begin{align}
V_{mm^{\prime}} = e^{-(\Delta m)^2 \Gamma L}
\end{align}
and for the N00N state
\begin{align}
V_{\textrm{N00N}} = e^{-N^2 \Gamma L}.
\end{align}
We note that the visibility of the N00N state with the parity detection in Eq.~(18) agrees with the visibility in Ref.~\cite{James}.

The visibility in Eqs.~(17) and (18) depends on the value of the dephasing rate $\Gamma$ and $N$ (or $\Delta m = m - m^\prime$), and for a given value of $\Gamma$, the visibility falls down faster as N increases. Hence, high-N00N states (or $mm^{\prime}$ states) with large number of photons are very much susceptible to the phase-fluctuations compared to the low-NOON states, and hence are not suitable to achieve metrological advantage with robustness in presence of phase noise. This is shown in Fig.~4, where we plotted the visibility for different N (or $\Delta m$) with respect to the dephasing rate $\Gamma$.

\begin{figure}[h!]
\centering
	\includegraphics[width=1\columnwidth,height=2.4in]{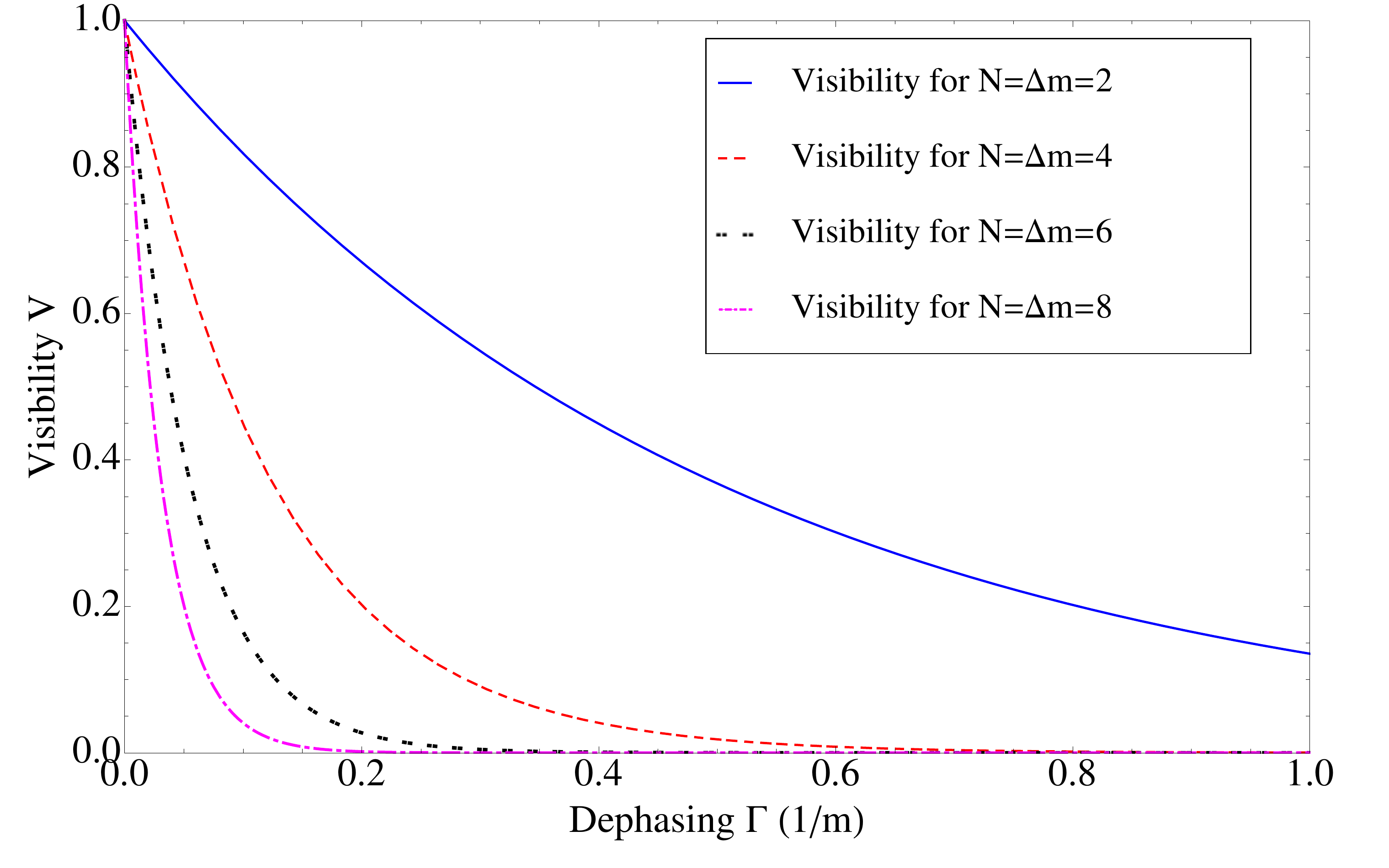}
	\caption{(Color online) Visibility $V$ of the $mm^{\prime}$ state for different $\Delta m$ (for different $N$ in case of N00N states with the same phase information) as a function of $\Gamma$. The visibility $V$ is plotted for $N$(or $\Delta m$)=2 [solid blue line], $N$(or $\Delta m$)=4 [dashed red line], $N$(or $\Delta m$)=6 [double dotted black line], and $N$(or $\Delta m$)=8 [dotdashed purple line). We see that the visibility drops faster for larger values of $\Delta m$ (or $N$).}
\end{figure}

\section{Quantum Fisher Information: Bounds for Phase sensitivity}
In order to minimize the uncertainty $\delta \phi$ of the measured phase, we now seek to provide the lowest bound on the uncertainty of the phase.
This bound is given by the quantum Cram\'er-Rao bound $\delta \phi_{\textrm{QCRB}} $, and is inversely proportional to the quantum Fisher information $F(\phi)$~\cite{Helstrom,Holevo, Braunstein1,Braunstein2}
\begin{align}
\delta \phi_{\textrm{QCRB}} \geq \frac{1}{\sqrt{F(\phi)}}
\end{align}
A general framework for estimating the ultimate precision limit in noisy quantum-enhanced metrology has been studied by Escher \emph{et al}~\cite{Davidovich}.
In the following, we first obtain the quantum Fisher information, leading to the quantum Cram\'er-Rao bound for both the $mm^{\prime}$ and N00N states in presence of the phase fluctuations, and show that the parity detection attains the quantum Cram\'er-Rao bound for both of these states subject to the dephasing.

The quantum Cram\'er-Rao bound has been shown to be always reached asymptotically by maximum likelihood estimations and a projective measurement in the eigenbasis of the symmetric logarithmic derivative $L_\phi$~\cite{PRL_Paris,Braunstein1,Braunstein2}, which is a self-adjoint operator satisfying the equation
\begin{align}
\frac{L_\phi \rho_\phi + \rho_\phi L_\phi}{2} = \frac{\partial \rho_\phi}{\partial\phi},
\end{align}
where $\rho_\phi$ is given by Eq. (6) for $mm^{\prime}$ state and by Eq. (7) for N00N state. The quantum Fisher information $F(\rho_\phi)$ is then expressed as~\cite{Paris}
\begin{align}
F(\rho_\phi)= \mathrm{Tr} (\rho_\phi L_\phi L_\phi^{\dagger}) = \mathrm{Tr}( \rho_\phi L_\phi^2).
\end{align}
The symmetric logarithmic operator $L_\phi$ is given by
\begin{align}
\frac{\lambda_i + \lambda_j}{2} \langle i|L_\phi|j\rangle = \langle i |\frac{\partial \rho_{\phi}}{\partial \phi}|j\rangle,
\end{align}
for all $i$ and $j$, where $\lambda_i$ and $|i\rangle$ are the eigenvalue and the corresponding eigenvector of $\rho_\phi$.
Evaluating $\rho_\phi$ and $\partial \rho_\phi/\partial \phi$ from Eq. (6) and then using Eqs. (21) and (22), we obtain the quantum Fisher information for the $mm^{\prime}$ state 
\begin{align}
F_{mm^\prime} = (\Delta m)^2 e^{-2(\Delta m)^2 \Gamma L}
\end{align}
leading to the quantum Cram\'er-Rao bound
\begin{align}
\delta \phi_{\textrm{QCRB}, mm^{\prime}} \geq \frac{1}{{\sqrt{F_{mm^{\prime}}}}} = \frac{1}{\Delta m e^{-(\Delta m)^2 \Gamma L}}.
\end{align}
For the N00N states, similar calculation with Eq.~(7) yields
\begin{align}
F_{\textrm{N00N}} = N^2e^{- 2 N^2 \Gamma L},
\end{align}
and
\begin{align}
\delta \phi_{\textrm{QCRB, N00N}} \geq \frac{1}{{\sqrt{F_{\textrm{N00N}}}}} = \frac{1}{N e^{-N^2 \Gamma L}}.
\end{align}
Eqs.~(24) and (26) represent the lowest bound on the uncertainty of the phase measurement for the $mm^{\prime}$ and N00N states, respectively.

For a detection scheme to be optimal, it has to saturate the quantum Cram\'er-Rao bound. Eqs.~(13) and (14) represent phase sensitivity for the $mm^{\prime}$ and N00N states respectively, and these expressions can be shown to be identical to the quantum Cram\'er-Rao bounds in Eqs. (24) and (26) for $\phi = \pi/(2 \Delta m)$, or $\phi = \pi/(2 N)$ for the $mm^{\prime}$ or N00N states respectively. Thus, parity detection saturates the quantum Cram\'er-Rao bounds and is optimal for both the states in presence of the phase fluctuations.


\section{Effects of both photon loss and phase fluctuations}

\subsection{Evolution}

Following Ref.~\cite{Kebei}, two fictitious beam splitters are added before stage I of our previous configuration to model photon loss from the system into the environment, as shown in Fig.~5. The two fictitious beam splitters have transmittance $T_a$ and $T_b$, and reflectance $R_a=1-T_a$ and $R_b=1-T_b$, respectively. General $T_a$ and $T_b$ are used in the following derivation of the density matrix, but later we assume $R_a=0$ to mimic the local path which is well-isolated from the environment. 
\begin{figure}[t!] 
\centering 
\includegraphics[width=\columnwidth]{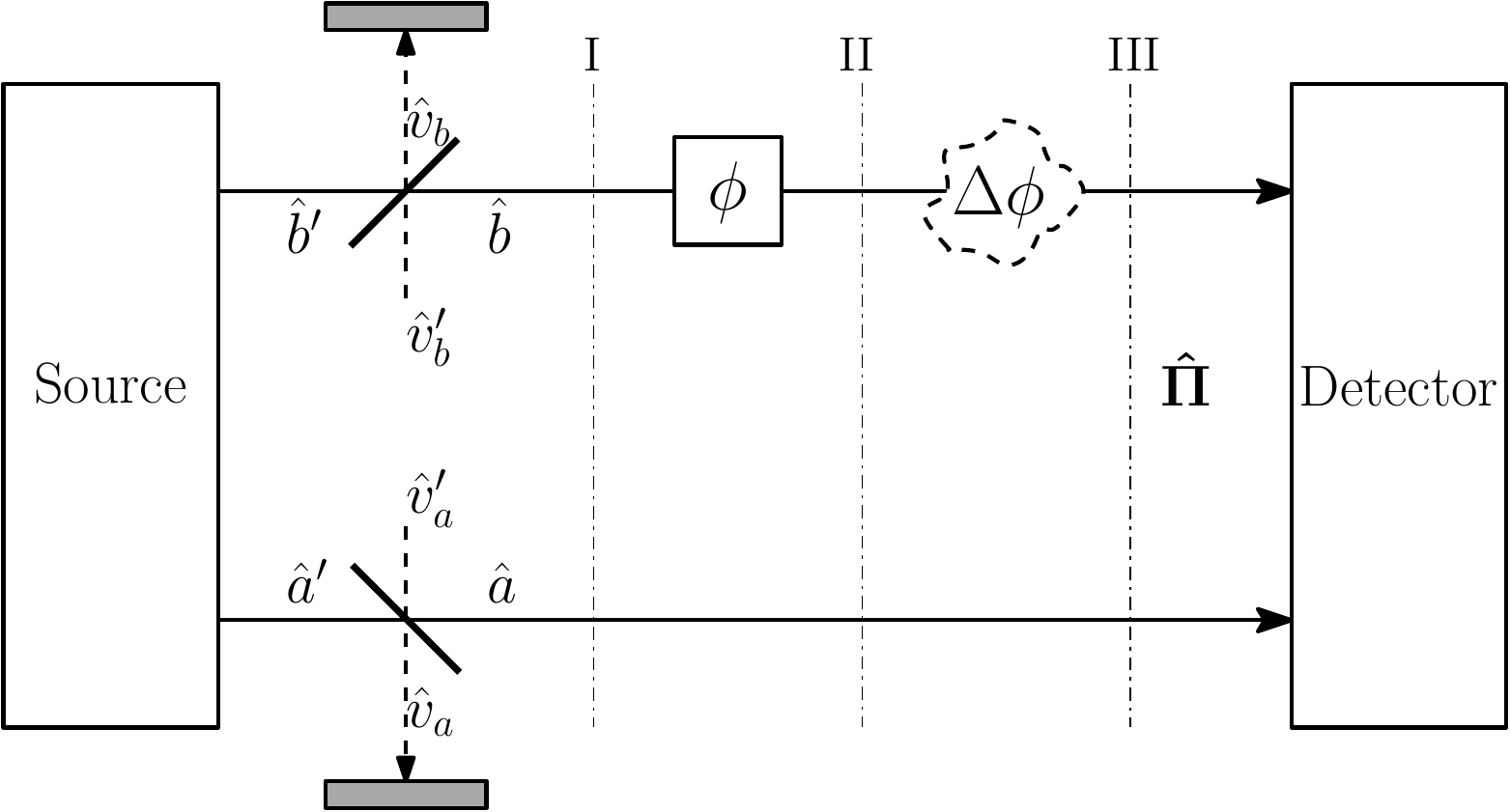} 
\caption{Two fictitious beam splitters are introduced to Fig.~1 to mimic the loss of photon from the system into the environment. After tracing out the environment mode $v_b$ and $v_a$, the system results in a mixed state at stage I. } 
\label{interferometer}
\end{figure}

The photon loss entangles the system with the environment and leaves the system in a mixed state. For a general $mm'$ input state, the density matrix of the system at stage II can be easily deducted from Ref.~\cite{Kebei} as
\begin{align} 
\nonumber
\rho_{mm'}(t) =&~\sum_{k=0}^{m} \sum_{k'=0}^{m'} \Bigg\lbrace \vert \alpha \vert^2 d_1(t) \vert k,k'\rangle \langle k,k' \vert  \\
\nonumber
& ~ \qquad  +\vert \beta \vert^2 d_2(t) \vert k',k\rangle \langle k',k \vert \Bigg\rbrace \\
\nonumber
& ~+\sum_{k=0}^{m'} \sum_{k'=0}^{m'} \Bigg\lbrace \alpha \beta^* d_3(t) \vert \Delta m +k,k'\rangle \langle k,\Delta m +k' \vert \\
& ~ \qquad + \alpha^* \beta d_4(t) \vert k',\Delta m +k\rangle \langle \Delta m +k',k \vert \Bigg\rbrace,
\label{eq:DM}
\end{align}
where $\alpha =e^{im'\phi}/\sqrt{2}$, $\beta =e^{im\phi}/\sqrt{2}$ as before, and the coefficients $d_i (i={1,2,3,4})$ are defined as
\begin{align} 
\nonumber
d_1(k,k',t=0)=&~  \binom{m}{k} \binom{m'}{k'} \vert T_a \vert^{k} \vert R_a \vert^{m-k} \vert T_b \vert^{k'} \vert R_b \vert^{m'-k'},\\
\nonumber
d_2(k,k',t=0)=&~  \binom{m}{k} \binom{m'}{k'} \vert T_a \vert^{k'} \vert R_a \vert^{m'-k'} \vert T_b \vert^{k} \vert R_b \vert^{m-k}, \\
\nonumber
d_3(k,k',t=0)=&~  \binom{m}{\Delta m +k}^{\frac{1}{2}} \binom{m}{\Delta m +k'}^{\frac{1}{2}} \binom{m'}{k}^{\frac{1}{2}} \binom{m'}{k'}^{\frac{1}{2}}\\
\nonumber
     &\times T_a^{\frac{1}{2}(\Delta m+2k)} R_a^{m'-k} T_b^{\frac{1}{2}(\Delta m+2k')} R_b^{m'-k'}, \\
\nonumber
d_4(k,k',t=0)=&~  \binom{m}{\Delta m +k}^{\frac{1}{2}} \binom{m}{\Delta m +k'}^{\frac{1}{2}} \binom{m'}{k}^{\frac{1}{2}} \binom{m'}{k'}^{\frac{1}{2}} \\
      &\times T_a^{\frac{1}{2}(\Delta m+2k')} R_a^{m'-k'} T_b^{\frac{1}{2}(\Delta m+2k)} R_b^{m'-k}.
      \label{eq:DM coefficients}
\end{align}
Given the system undergoes pure dephasing after stage II, we may use previous result and show that
the evolution of the density matrix $\rho_{mm'}(t)$ is
\begin{align}
\nonumber
\dot{\rho}_{mm'}(t)=&-{\Delta m}^2 \Gamma \times \\
\nonumber
&\sum_{k,k'=0}^{m'} \Bigg\lbrace \alpha \beta^* d_3(t) \vert \Delta m + k,k'\rangle \langle k,\Delta m +k' \vert \\
~& \qquad ~ + \alpha^* \beta d_4(t) \vert k',\Delta m +k\rangle \langle \Delta m +k',k \vert \Bigg\rbrace.
\end{align}
It is then easy to see that $d_{1}(t)=d_{1}(0)$, $d_{2}(t)=d_{2}(0)$, $d_{3}(t)=e^{-{\Delta m}^2 \Gamma L} d_{3}(0)$ and $d_{4}(t)=e^{-{\Delta m}^2 \Gamma L} d_{4}(0)$.

\subsection{Phase-Sensitivity and Visibility}
Similar to Ref.~\cite{Kebei}, we define
\begin{align}
\nonumber
K_1(t)=\displaystyle\sum_{k=0}^{m'}\left( d_1(k,k,t)+d_2(k,k,t) \right), \\
K_2(t)=\displaystyle\sum_{k=0}^{m'}\left( d_3(k,k,t)+d_4(k,k,t) \right),
\end{align}
and it is straightforward to show that $K_1(t)=K_1(0)$ and $K_2(t)=K_2(0)e^{-\Delta m^2 \Gamma L}$. From Eqs.~(10) and (27), the parity signal of a $mm'$ state under both photon loss and phase fluctuation can be shown to be
\begin{align}
\langle \hat{\Pi}_{mm'} \rangle=K_1(t)+(-1)^{m+m'}K_2(t)\cos(\Delta m \phi). 
\end{align}
This gives rise to the phase-sensitivity for the parity detection for a $mm'$ state under both photon-loss and phase-fluctuations as
\begin{align}
\delta \phi_{mm'}=\sqrt{\frac{1-\left\lbrace K_1(t)+(-1)^{m+m'}K_2(t)\cos(\Delta m \phi) \right\rbrace^2}{ \left\lbrace \Delta m K_2(t) \sin(\Delta m \phi) \right\rbrace^2 }},
\end{align}
where linear error propagation method in Eq.~(12) is employed. Notice that when loss is negligible this sensitivity recovers Eq.~(13). 

A relative visibility with respect to both loss and phase-fluctuations can be defined as
\begin{align}
\nonumber
V_{mm^\prime} &= \frac{\langle \hat{\Pi}_{mm'}\rangle_{\textrm{max}} - \langle \hat{\Pi}_{mm^\prime}\rangle_{\textrm{min}}}{\langle \hat{\Pi}_{mm^\prime}(\Gamma=0, L=0)\rangle_{\textrm{max}} - \langle \hat{\Pi}_{mm^\prime}(\Gamma=0, L=0)\rangle_{\textrm{min}}}, \\
& = K_2(0) e^{-\Delta m^2 \Gamma L}
\end{align}
where $L=R_b$ characterizes the loss in the upper path and $R_a$ is set to be zero as aforementioned. In the limit of $L\rightarrow0 $, $K_2(0)$ approaches one and the visibility reduces to the previous result. Notice the dephasing only affects the off-diagonal terms of the density matrix while loss affects both diagonal and off-diagonal terms. However, because of the linearity of the Mach-Zehnder interferometer, the effect from photon loss is independent of that from phase-fluctuation, as expected. All results in this section apply to N00N states with $N=m$ and $m'=0$. 


\section{Summary}

In this work, we studied the effects of phase fluctuations on the phase sensitivity and visibility of $mm^{\prime}$ and N00N states 
in an optical interferometric setup. Although $mm^{\prime}$ states are more robust than N00N states against photon loss, we showed that 
they do not provide any better performance in presence of such phase fluctuations than their equivalent N00N counterpart. We have used
the parity detection technique for phase estimation, that can be readily implemented using photon-number-resolving detectors~\cite{Christoph}
in the low power regime and using optical nonlinearities and homodyning in the high power regime~\cite{Gerry2,Gerry-Bui,Plick,Jiang}. Using the same detection technique, we explicitly derived the phase-sensitivity and the visibility in a more general case with both the photon loss and phase-fluctuations. We have also
presented a brief study on the quantum Fisher information for both the $mm^{\prime}$ and N00N states and showed that the parity detection serves as 
the optimal detection strategy in both cases as it saturates the quantum Cram\'er-Rao bound of the interferometric scheme.

\section{Acknowledgment}

We would like to acknowledge helpful discussions with Daniel James and Luiz Davidovich. This work is supported by the grants from the Air Force Office of Scientific Research and the National Science Foundation.

\end{document}